\documentclass[12pt, draftclsnofoot, onecolumn]{IEEEtran}
\IEEEoverridecommandlockouts
\usepackage[noadjust]{cite}
\usepackage[numbers,sort&compress]{natbib}
\usepackage{amsmath,amssymb,amsfonts}
\usepackage{algorithmic}
\usepackage{graphicx}
\usepackage{textcomp}

\usepackage{makecell}

\usepackage[parfill]{parskip}

\usepackage[switch, displaymath, mathlines]{lineno}

\usepackage{graphicx}
\usepackage{epstopdf}
\usepackage{epsfig}
\usepackage{subfig}

\usepackage[table]{xcolor}
\definecolor{lightgray}{gray}{0.9}

\usepackage{etoolbox}
\makeatletter
\patchcmd{\@makecaption}
  {\scshape}
  {}
  {}
  {}
\makeatletter
\patchcmd{\@makecaption}
  {\\}
  {.\ }
  {}
  {}
\makeatother

\usepackage{mathtools}
\usepackage{abraces}

\usepackage[breaklinks=true]{hyperref}
\hypersetup{
    colorlinks=true,       
    linkcolor=blue,          
    citecolor=blue,        
    filecolor=magenta,      
    urlcolor=cyan           
}

\makeindex

\begin{document}

\title{\Huge  Deep-Learning based Multiuser Detection for NOMA} 

\author{Krishna Chitti, Joao Vieira, Behrooz Makki,~\IEEEmembership{Senior Member,~IEEE} \\
Ericsson AB, Sweden \\
firstname.lastname@ericsson.com
}

\maketitle
\begin{abstract}
In this paper, we study an application of deep learning to uplink multiuser detection (MUD) for non-orthogonal multiple access (NOMA) scheme based on Welch bound equality spread multiple access (WSMA).
Several non-cooperating users, each with its own preassigned NOMA signature sequence (SS), transmit over the same resource.
These SSs have low correlation among them and aid in the user separation at the receiver during MUD.
Several subtasks such as equalizing, combining, slicing, signal reconstruction and interference cancellation are involved in MUD.
The neural network (NN) considered in this paper replaces these well-defined receiver blocks with a single black box, i.e., the NN provides a one-shot approximation for these modules.
We consider two different supervised feed-forward NN implementations, namely, a deep NN and a $2$D-Convolutional NN, for MUD.
Performance of these two NNs is compared with the conventional receivers.
Simulation results show that by proper selection of the NN parameters, it is possible for the black box approximation to provide faster and better performance, compared to conventional MUD schemes, and it achieves almost the same symbol error rate as the ultimate one obtained by the complex maximum likelihood-based detectors.
\end{abstract}

\begin{IEEEkeywords}
NOMA, WSMA, multiuser detection, supervised learning, deep neural network, convolutional neural network, machine learning, multiuser detection, deep learning
\end{IEEEkeywords}

\section{Introduction}
\label{sec:intro}
In the last decade, non-orthogonal multiple access (NOMA) has received considerable attention as a candidate multiple access (MA) technique for certain scenarios, such as asynchronous access in massive machine type communications \cite{makki2020survey}.
With NOMA, multiple user equipments (UEs) share the same radio resources in time, frequency or code.
Different fundamental results have been derived for the performance analysis of NOMA in downlink \cite{sun2017optimal, mokhtari2018download, xu2015noma, yang2016general, ding2016general} and uplink (UL) \cite{ yang2016general, ding2016general, zhang2016uplink, tabassum2017modeling}.
As shown in these works, with appropriate parameter settings, NOMA can potentially outperform existing orthogonal MA schemes at the cost of, e.g., advanced receiver, UE pairing and coordination complexity.

In 3GPP, NOMA was first introduced as an extension of network-assisted interference cancellation and suppression for inter-cell interference mitigation in LTE Release 12 \cite{3GPPTR36866V1201}, as well as a study-item of LTE Release 13, under the name of downlink multi-user superposition transmission \cite{3GPPRP150496}. 
Additionally, the performance of various NOMA schemes and their implementation challenges were investigated as a part of Release 15 study-item on NOMA \cite{makki2020survey, 3GPPTR38812}. 

From another perspective, high user density, low latency requirements with high reliability and multi-antenna techniques have increased the overall complexity of the transmitter-receiver chain.
To address these, neural networks (NNs) are currently considered as one of the enablers to reduce the complexity and generate faster outputs.
The NN could be a partial substitute to some of the blocks in the communication chain, particularly when the channel statistics are not severely varying.
This is important in NOMA setups, since the presence of multiple non-orthogonal signals and their mutual interference may make the implementation of these modules challenging.
With an increase in computational power and a better data representation, the NN's performance is improving and reaching the one obtained by the conventional methods.

A well designed transmitter or receiver chain has several concatenated blocks, each with a predefined and optimized functionality.
With the NN implementation, few blocks may be replaced by a single or multiple black boxes, that learn and implement the optimized functions, i.e., the NN approximates the replaced modules.
The approximation mainly consists of linear operations followed by element wise nonlinear operations, leading to a simpler implementation with low latency.
Also, with an NN-based solution, the need for analytic description is eliminated, and it is easier to scale the network as the number of UEs or the transmission conditions change.

Deep learning (DL) for downlink multiuser multi-antenna NOMA is studied in \cite{kang2019deep}, where the transmitter precoding and the receiver decoding are jointly learned.
The NOMA variant considered in \cite{kang2019deep} is the conventional power-domain NOMA \cite{saito2013non}, where transmitter side superposition coding and receiver side successive interference cancellation (SIC) are adopted.
The total mean squared error (MSE) and the per UE bit error rate (BER) were simultaneously improved by the DNN implementation for a two UE case.
Similarly in \cite{lin2019deep}, a two user power-domain NOMA is considered where a significant symbol error rate (SER) performance improvement of a deep neural network (DNN) receiver over a conventional SIC is observed.
To handle channel distortions in various conditions, a DL receiver is considered in \cite{ye2017initial} for a single UE.
The receiver side NN facilitates simultaneous channel equalization and decoding to improve the bit error rate (BER).
With an inclusion of orthogonal frequency division multiplexing (OFDM) at the transmitter, \cite{ye2017power} extends the results of \cite{ye2017initial}, and  shows an improvement of the NN's BER performance despite dropping the cyclic prefix and introducing signal clipping. 
The design of DL methods for joint multiple-input multiple-output (MIMO) detection and channel decoding for various single UE MIMO configurations is studied in \cite{wang2019deep}, where, compared to the conventional methods, an improved BER performance is achieve by the trained NN. 
A concatenation of convolutional NN (CNN) based channel equalizer and a DNN based decoder is considered in \cite{xu2018joint} to improve the single UE BER performance.
Also, \cite{baek2019implementation} considers a classification-based DNN for single UE detection, where the channel estimation and detection is a one-shot implementation, and the uncoded BER performance of the DNN matches the maximum-likelihood (ML) detector performance.

In this paper, we study the application of a feed-forward supervised NN for symbol-level MUD of UL Welch bound equality
spread multiple access (WSMA) NOMA transmission.
WSMA is a transmitter side scheme where low-correlation spreading vectors, one each at the simultaneously transmitting UEs, are used to achieve UE separation at the receiver. 
To design the proposed MUD scheme, we consider two different NNs, a fully-connected NN and a $2$D-convolutional NN, and train them separately.
The training is followed by a test phase, where the NNs replace the conventional model based receiver-side blocks by a single black box for providing a one-shot approximation of the involved MUD subtasks.
These subtasks include equalizing, combining, interference cancellation, slicing and signal reconstruction. 
The NN's performance is studied over a broad range of signal-to-noise ratios (SNRs), both for varying number of UEs and different labelling formats at training, and we compare the performance of our proposed schemes with those in state-of-the-art MUD techniques.

We show through simulations, that with sufficient training and proper preprocessing of the NN's input, each of the considered NNs provides a better SER performance, compared to the well-known conventional non-linear detectors techniques.
Also, at low/moderate SNRs, the NN-based schemes achieve almost the same SER as the ultimate one obtained by the complex ML-based detectors. 
Finally, this performance is obtained with low latency since the involved calculations are simple mathematical operations.

The outline of this paper is as follows.
Section \ref{subsec:sys_model} explains the system model of the considered NOMA based UL multiuser single-input multiple-output (MU-SIMO) setup.
Spreading vectors for WSMA-NOMA are explained in Section \ref{subsubsec:sig_seq}.
Section \ref{sec:dnn_mud} gives the details of the NN (that operates as a classifier) based MUD along with the definitions of various NN parameters and functions.
For supervised learning, the NN's training data requires labelling.
Two separate labelling formats are considered here, whose description is given in Section \ref{subsec:classifier_labelling}.
The NN's architecture details are provided in \ref{subsec:nn_arch}.
Two different architectures are outlined here, a fully-connected NN given in Section \ref{subsubsec:fcnn} and a CNN given in \ref{subsubsec:cnn}.
Simulation results are discussed in Section \ref{sec:results}, and Section \ref{sec:conclusion} concludes the paper.

\section{WSMA based NOMA}
\label{sec:wsma_noma}
With an aim to improve spectral efficiency while supporting a higher UE density, NOMA was explored as one of the candidate technologies for 5G in 3GPP Rel-16  \cite{R11802767}. 
Several forms of NOMA exist which mainly differ in the way non-orthogonality is implemented, e.g., bit-level and symbol-level, and the domain, e.g., power, time and frequency, in which it is introduced.
See \cite[Fig 1, Table I]{dai2018survey} for a summarization of various NOMA schemes.
A low correlation symbol-level frequency domain spreading scheme, known as WSMA-NOMA, is considered here.
This is explained next.

\subsection{System Model}
\label{subsec:sys_model}
\begin{figure}[!t]
\centering
\includegraphics[]{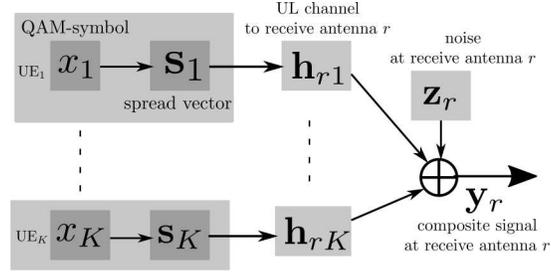} 
\caption{Block diagram for UL NOMA transmission.}
\label{fig:tx_block}
\end{figure}

Fig. \ref{fig:tx_block} shows the block diagram of a transmitter-side NOMA implementation. 
It's mathematical representation is given in \eqref{eqn:01}  and  \eqref{eqn:02}.
From Fig. \ref{fig:tx_block} it is seen that each of the $K$ UEs modulates its transmit symbol by a preassigned spread-vector before transmitting it over the shared UL channel.
At the access point (AP), the received signal is the noisy version of the sum of all the UEs transmit vectors, which occupy the same set of resource elements (REs).
With $\odot$ denoting the element-wise product, the $(L\times 1)$ UL complex received baseband sum-signal vector at the AP's receive antenna $r$ is
\begin{equation}
\label{eqn:01}
\mathbf{y}_r = \sum\limits_{k=1}^K \;(\mathbf{h}_{kr} \odot \mathbf{s}_k) \, x_k  + \mathbf{z}_r,
\end{equation}
where for a UE$_k$,  $x_k$ is the transmit symbol, $\mathbf{h}_{kr}$ is the $(L\times 1)$ fading channel and $\mathbf{s}_k$ denotes the $(L\times 1)$ preassigned spread-vector or signature sequence (SS).
The parameter $L$ is known as the spreading length of the SS.
The $(L\times 1)$ zero-mean additive white Gaussian noise (AWGN) vector at receive antenna $r$ is $\mathbf{z}_r \sim \mathcal{CN}(\boldsymbol 0,\sigma_{{\mathbf{z}}}^2\mathbf{I}_L ), \, \forall r$, where  $ \sigma_{{\mathbf{z}}}^2/2$ denotes the noise variance per complex dimension. 
The transmit symbols $x_k\in\mathcal{Q}, \, \forall k$, where $\mathcal{Q}=\{ q_1,q_2,\cdots,q_M \}$  is the set of $M$ QAM-symbols.
The UL channels $\mathbf{h}_{kr}\sim \mathcal{CN}(\boldsymbol 0, \sigma_{\text{h}}^2 \,\mathbf{I}_{L}), \forall k, r$, are i.i.d, where the channel variance per complex dimension is $\sigma_{\text{h}}^2/2$ and $\mathbf{I}_L$ is the $(L \times L)$ Identity matrix.

The overall setup in \eqref{eqn:01} can be rewritten as 
\begin{equation}
\label{eqn:02}
\mathbf{y} = \mathbf{H}_{\textrm{eff}}\mathbf{x} + \mathbf{z},
\end{equation}
where $\mathbf{y}=[\mathbf{y}_{1}^H, \mathbf{y}_{2}^H, \cdots, \mathbf{y}_{N_{\text{r}}}^H]^H$ represents the $(L N_{\text{r}} \times 1)$ joint sum-signal of the $K$ UEs at the AP,  $\mathbf{x}=[x_1, x_2, \cdots, x_K]^T$ is the $(K \times 1)$ transmit vector of QAM-symbols from the $K$ UEs, $\mathbf{z}=[\mathbf{z}^H_1, \mathbf{z}^H_2, \cdots, \mathbf{z}^H_{N_{\text{r}}}]^H$ is the $(L N_{\text{r}} \times 1)$ noise vector at all $N_{\text{r}}$ receive antennas.
Also, $\mathbf{H}_{\textrm{eff}}=[\mathbf{h}_{1,\textrm{eff}}, \mathbf{h}_{2,\textrm{eff}}, \cdots, \mathbf{h}_{K,\textrm{eff}}]$ is the $(L N_{\text{r}} \times K)$ effective UL channel matrix from all $K$ UEs to the AP.
$(L N_{\text{r}} \times 1)$ vector $\mathbf{h}_{k,\textrm{eff}}=[\mathbf{h}^H_{k1,\textrm{eff}}, \mathbf{h}^H_{k2, \textrm{eff}}, \cdots, \mathbf{h}^H_{kN_{\text{r}},\textrm{eff}}]^H$, and $(L \times 1)$ vector $\mathbf{h}_{kr, \textrm{eff}}=(\mathbf{h}_{kr} \odot \mathbf{s}_k)$ denote the effective UL channels from UE$_k$ to all $N_{\textrm{r}}$ receive antennas and to a receive antenna $r$ respectively.
The transpose and the Hermitian operations are denoted as  $(\cdot)^T$ and $(\cdot)^H$ respectively.

\subsubsection{Signature Sequences} 
\label{subsubsec:sig_seq}
The spread-vectors $\mathbf{s}_k,\forall k$, in \eqref{eqn:01} repeat the QAM-symbols at each UE over $L$ REs in a weighted manner.
If frequency domain spreading is considered, $\mathbf{y}$ in \eqref{eqn:02} is a joint space-frequency vector.
The scalar value $\zeta=(K/L)$, known as the overloading factor, indicates the UE density per RE.
For WSMA, it is required to have $K\geq L$.
At the AP, the $(L\times K)$ spread matrix $\mathbf{S}=[\mathbf{s}_1,\mathbf{s}_2,\cdots,\mathbf{s}_K]$ is pregenerated, and the SSs are preassigned to the $K$ UEs in a mutually exclusive manner to avoid SS collisions.
With  $|\cdot|$ denoting the absolute value, the correlation among the SSs is given as 
\begin{equation}
  \rho_{kj} = |\mathbf{s}_k^H\mathbf{s}_j| =
    \begin{cases}
      < 1, & \text{if } k\neq j,\\
      = 1, & \text{if } k=j.\\
    \end{cases}       
\end{equation}

For the spread-matrix $\mathbf{S}$ generation with a given correlation property, a scalar performance indicator (PI) is optimized.
Two such PIs are the total squared correlation, (TSC), \cite[Equation 1.32]{popescu2006interference} and the worst case matrix coherence, $\mu$, \cite[Equations 1, 6]{calderbank2015block} which are, respectively, given as
\begin{align}
\textrm{TSC} & = \sum\limits_{k=1}^{K}\sum\limits_{j=1}^{K}  \rho_{kj}^2, \label{eqn:PI_tsc} \\
 \textrm{and}\quad \mu & = \underset{\forall k,\forall j; \; k \neq j}{\text{maximum}}\; \rho_{kj}. \label{eqn:PI_mu}
\end{align}
Optimizing the PIs in \eqref{eqn:PI_tsc} and \eqref{eqn:PI_mu} separately will result in an $\mathbf{S}$ with different properties.
For better signal separability at the AP, it is required to have small values for $\rho_{kj}$, hence \eqref{eqn:PI_tsc} and \eqref{eqn:PI_mu} are minimized for lower values of TSC and $\mu$.
With WSMA-NOMA, in addition to the spatial separation due to the presence of multiple antennas, frequency domain separation is also achieved, thus aiding the MUD of interfering UEs.

There exists a lower bound, known as the Welch bound (WB), on the TSC, such that $(K^2/L)<=\textrm{TSC}$.
The spread-matrix $\mathbf{S}$ at the optimal value $K^2/L=\textrm{TSC}$ during minimization is called the Welch bound equality (WBE) set, hence the name WSMA \cite{R11802767}.
With optimizing the TSC, the vectors of $\mathbf{S}$ converge as an ensemble.
Thus the spread-matrix $\mathbf{S}$ is not unique and there exist many $\mathbf{S}$ for the same TSC.
The term WBE-set in general encompasses several WBE-subsets, each with its own correlation properties.
One such WBE subset is the line-packing Grassmann set \cite{medra2014flexible}, obtained from \eqref{eqn:PI_mu}.
The optimal solution of $\mu$ will result in an $\mathbf{S}$ such that every two SSs in $\mathbf{S}$ have the same correlation value, i.e., an equiangular property with $\rho_{kj}=\rho,\forall k \neq j$. 

\subsection{Multiuser Detection}
\label{subsec:mud}
A symbol-level multiuser detector finds an estimate $\hat{\mathbf{x}} $ for a transmitted symbol $\mathbf{x}$ from the received sum-signal $\mathbf{y}$.
For MUD, several subtasks such as equalizing, combining, interference cancellation (IC), slicing and signal reconstruction are carried out.
Information in the form of estimates of both the UE channels and interfering signals' for some or all the UEs is made available for MUD.
With equi-probable transmission of the QAM-symbols and AWGN assumption at the AP, the optimal detector is the maximum-likelihood (ML) detector, whose estimate is obtained as 
\begin{equation}
\label{eqn:03}
\hat{\mathbf{x}} = \underset{\bar{\mathbf{x}}_n \in \mathcal{D}_{\mathbf{x}}, \, \forall n }{\text{arg-min}}\; \| \mathbf{y} - \hat{\mathbf{H}} \bar{\mathbf{x}}_n\|^2_2,
\end{equation} 
where  $\|\cdot\|_2$ is the $\ell_2$-norm, $\hat{\mathbf{H}}=[\hat{\mathbf{h}}_1, \hat{\mathbf{h}}_2, \cdots, \hat{\mathbf{h}}_K]$ is the  $(L N_{\text{r}} \times K)$ overall channel estimate at the AP and $\hat{\mathbf{h}}_k$  represents the $(L N_{\text{r}} \times 1)$ estimate of the effective channel at all $N_{\text{r}}$ antennas for a UE$_k$.

Solving  \eqref{eqn:03} involves an exhaustive search over $\mathcal{D}_{\mathbf{x}}=[\mathbf{x}_1,\mathbf{x}_2,\cdots,\mathbf{x}_{M^K}]$, which is the set of all possible QAM-symbol combinations of the $K$ UEs, and may be difficult to implement with increasing number of parameters.
As shown in \cite{moshavi1996multi}, to reduce the complexity, suboptimal linear techniques based on joint minimum mean squared error (MMSE) and matched filter (MF) exist.
The MF excludes the interfering UE channel estimates while detecting symbols for each UE, leading to a higher performance loss compared to the other linear techniques.
The joint-MMSE technique, depending on the operating point, is also interference-limited.
To further enhance the detector's performance, non-linear IC techniques, such as parallel-IC (PIC) or sequential-IC (SIC) may be implemented \cite{moshavi1996multi}.
The IC techniques primarily suffer from error propagation, where a detection error for a UE may lead to an error in the following detection steps.
For the SIC, an implicit problem of UE ordering also exists, while for the PIC the quality of the solution depends on the number of times IC is performed at each UE.
Thus, the latency in obtaining a solution increases with both the increasing number of UEs and the increasing IC stages.
For these reasons, depnding on the network size, an NN may be substituted to address both the latency and implementation complexity requirements while generating a an appropriate suboptimal solution.

\section{Deep Learning based multiuser detector}
\label{sec:dnn_mud}
\begin{figure}[!t]
\centering
\includegraphics[]{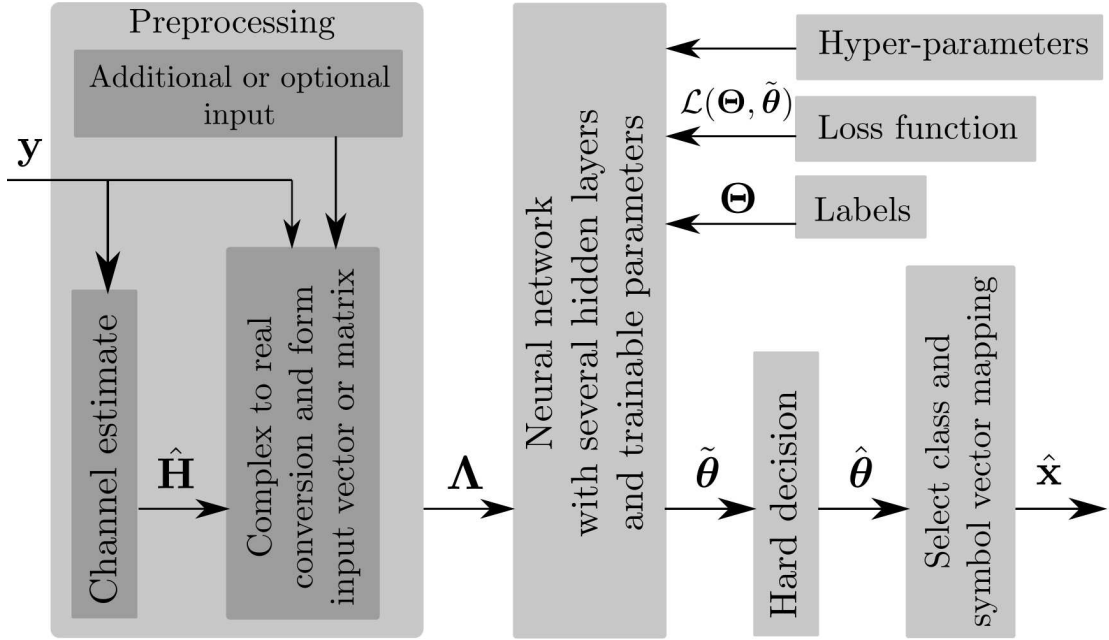} 
\caption{Block diagram of an NN-based classifier for symbol level MUD.}
\label{fig:nn_block}
\end{figure}

Block diagram of a supervised NN for MUD is shown Fig. \ref{fig:nn_block}, whose operation is discussed in this section.
It has two separate operating phases; a training phase followed by an testing phase.
The aim of training is to optimize and prepare the NN for the test phase.
It is in the test phase that the NN's functionality as a detector is realized.
In both phases, the received signal $\mathbf{y}$ and the channel estimates $\hat{\mathbf{H}}$ are preprocessed before feeding them as an input to the NN.
Preprocessing here involves reshaping and concatenating $\mathbf{y}$ and $\hat{\mathbf{H}}$ in an appropriate way to fit the NN's input dimension (a vector or a matrix).
The NN considered here is assumed to handle real valued inputs.
Thus, changing complex valued numbers to real valued is the last step of preprocessing.
Let $\boldsymbol\Lambda$ be the real valued input of appropriate dimension to the NN.
Additional information that probably may aid the detection can be included into $\boldsymbol\Lambda$, e.g., the spread-matrix $\mathbf{S}$ when $\hat{\mathbf{H}}$ is the estimate for only the propagation channel $\mathbf{H}$ and not the effective channel $\mathbf{H}_{\textrm{eff}}$. 


The NN has several hidden layers with trainable parameters, called weights and biases, that are optimized at training. 
\begin{figure}[!t]
\centering
\includegraphics[]{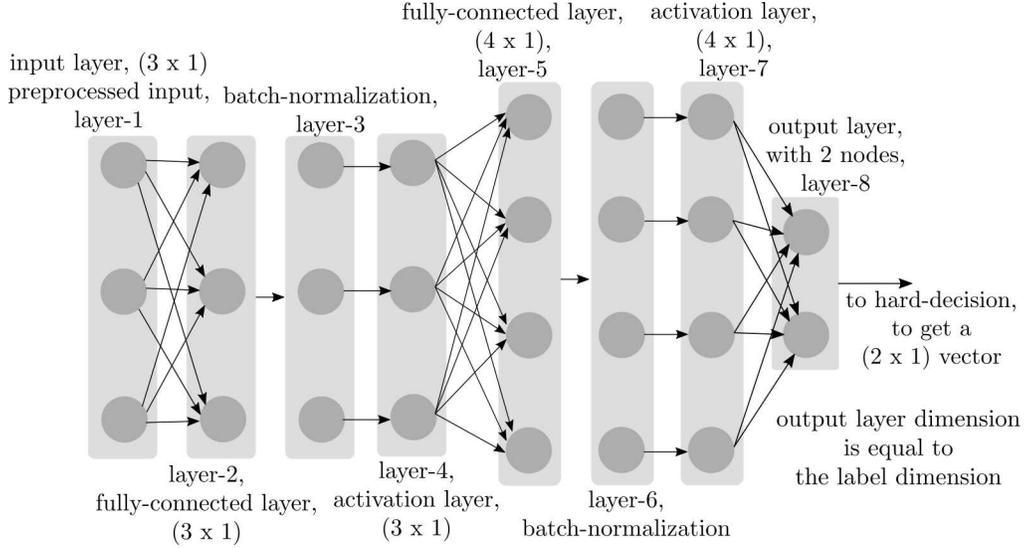} 
\caption{An example of $8$-layer fully-connected NN.}
\label{fig:fcnn_exm}
\end{figure}
As seen from an example in Fig. \ref{fig:fcnn_exm}, there are multiple layers with varying number of nodes per layer and forward connections between the nodes of different layers.
How the layers are formed, arranged and connected to form a structure is the NN's architecture.
The input passes sequentially through a fully-connected layer, a batch-normalization (BN) layer and an activation layer.
The BN layer ensures that the data while passing through the NN is properly scaled.
The activation layer introduces non-linearity and has no trainable parameters.
Commonly used activation function for the hidden layers is the Rectified Linear Unit (ReLU), which for an input $\mathbf{d}$, generates an element-wise output as $\textrm{max}(0,d_i)$, where $d_i$ is the $i^{\textrm{th}}$ component of $\mathbf{d}$ and the $\textrm{max}(\cdot, \cdot)$ function outputs the maximum of its two inputs.
Several such fully-connected, BN, ReLU activation layers are sequentially repeated before reaching the output.
For the output layer of a classifier, either a softmax or a sigmoid activation is selected.
Also, the number of output layer nodes is determined by the NN's functionality.
These are explained in Section \ref{subsec:classifier_labelling}.
 
The dataset here is a union of the set $\mathcal{D}_{\mathbf{x}}$, the channel samples $\mathbf{H}$ (the channel estimates $\hat{\mathbf{H}}$ to be precise) and the noise samples $\mathbf{z}$.
Since only the statistics of $\mathbf{H}$ and $\mathbf{z}$ are available, there are infinitely many realizations of $\mathbf{H}$ and $\mathbf{z}$ in the dataset.
Nevertheless, the term dataset here is used for the set $\mathcal{D}_{\mathbf{x}}$ for different values of $\mathbf{H}$ and $\mathbf{z}$.
At training, the dataset is sent through the NN multiple times.
The term epoch is used to indicate that the set $\mathcal{D}_{\mathbf{x}}$ was sent through the NN once.
Let the number of epochs be given as $N_{\textrm{epoch}}$.
The dataset is divided into several smaller subsets to avoid memory overflow issues on the machine that trains the NN.
Let $N_{\textrm{batch}}$, called the batch-size, be the size of every subset.

A supervised NN requires labelling of it's finite dataset $\mathcal{D}_{\mathbf{x}}$.
For this, every QAM-symbol combination $\mathbf{x}_n \in \mathcal{D}_{\mathbf{x}}$, is assigned a unique $\{0,1\}$ binary valued label vector $\boldsymbol\theta_n$.
Let $\boldsymbol\Theta=[\boldsymbol{\theta}_1, \boldsymbol{\theta}_2, \cdots, \boldsymbol{\theta}_{M^{K}}]$, be the set of all label vectors and $N_{\textrm{label}}$ be the length of every label vector in $\boldsymbol\Theta$.
Further details of the labelling are given in Section \ref{subsec:classifier_labelling}.

For an input $\boldsymbol\Lambda$, the NN generates a real and positive valued output vector $\tilde{\boldsymbol{\theta}}$ of length $N_{\textrm{label}}$, as a function of it's hidden layers.
By comparing the NN's output $\tilde{\boldsymbol{\theta}}$ to every member of the label set $\boldsymbol\Theta$, a scalar metric, called the loss function, is computed at training. 
The overall loss over one epoch, given as the sum of individual loss functions, is the measure of accuracy between the NN's inputs and the NN's outputs.
Let $\mathcal{L}(\boldsymbol\Theta,\tilde{\boldsymbol\theta}) = \sum\limits_{n=1}^{M^K}\mathcal{L}(\boldsymbol\theta_n,\tilde{\boldsymbol\theta})$, be the overall loss function and $\mathcal{L}(\boldsymbol\theta_n,\tilde{\boldsymbol\theta})$ be the loss function for a single sample in the dataset.
During the $N_{\textrm{epoch}}$ iterations, i.e., when the NN's tunable parameters are being optimized, the loss $\mathcal{L}(\boldsymbol\Theta,\tilde{\boldsymbol\theta})$ is minimized by using a gradient descent (GD) based algorithm \cite{ruder2016overview}.
The rate at which the GD algorithm converges is governed by the step size it uses to update it's intermediate solutions. 
Let $\alpha$ be the step size, called the NN's learning rate, of the GD algorithm.

The number of epochs $N_{\textrm{epoch}}$, the batch-size $N_{\textrm{batch}}$, the learning rate $\alpha$, are called the hyper-parameters.
The quality of test phase solution and the training phase convergence depend on these hyper-parameters.
In general, finding a good combination of all the hyper-parameters can be difficult.
In our simulations, we have verified the results for the cases with a few different values of hyper-parameters and compared the performance with the ML. 

\begin{figure}[!t]
    \centering
        \subfloat[Multi-Class with $K$ symbols per class. \label{fig:nn_output_e_left}]{\includegraphics[width=0.35\columnwidth]{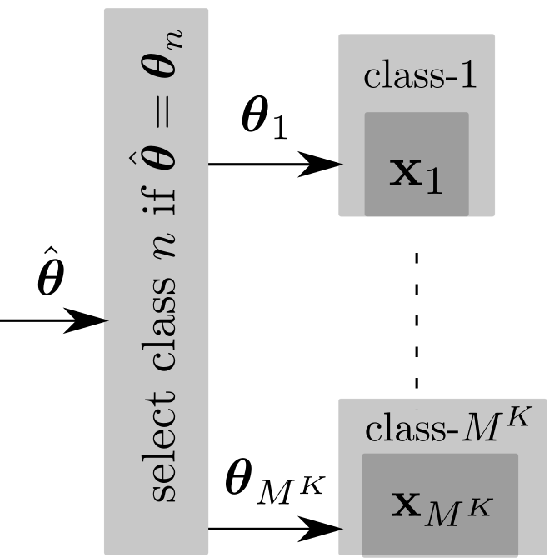}}
        \quad
        \subfloat[Multi-Label with one symbol per class.\label{fig:nn_output_e_right}]{\includegraphics[width=0.61\columnwidth]{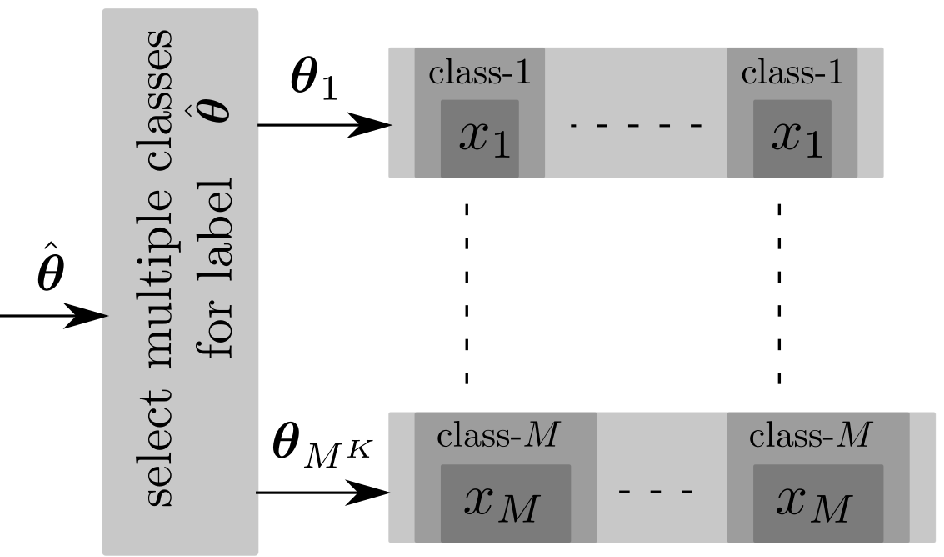}}
        \caption{Two possible outputs (labelling) for a classifier.}
         \label{fig:nn_output}
\end{figure}

After the offline training, the NN is ready for the test phase where no optimization is involved.
The NN's output $\tilde{\boldsymbol\theta}$ is sent for a hard-decision to generate a $\{0,1\}$ binary valued label estimate $\hat{\boldsymbol\theta}\in \boldsymbol\Theta$, which in turn points to the probably transmitted $K$ QAM-symbols.
Thus, the NN along with a hard-decision predicts the MUD's solution estimate $\hat{\mathbf{x}}$ from it's input $\boldsymbol\Lambda$.
The test phase prediction can be seen as a classification operation, which means that the NN's input is classified into one of the possible classes (see Fig. \ref{fig:nn_output}), where a class can be a member of the QAM-alphabet $\mathcal{Q}$ or the QAM-symbol set $\mathcal{D}_{\mathbf{x}}$.
This classification is explained in the following sections with examples in Tables \ref{tab:softmax_def}  and \ref{tab:sigmoid_def}.
These examples are given for $K=2$, $M=4$ and $\mathcal{Q}=[q_1,q_2,q_3,q_4]$.  
Also, only the first $4$ and the last $4$ entries of the possible $4^2$ entries are shown.

\subsection{Classifier and Labelling}
\label{subsec:classifier_labelling}
For the NN to operate as a classifier, the loss function, the hard-decision function, the number of nodes, the activation function of the output layer and the length of every label vector are all jointly selected.
Two types of classifiers and their associated labelling considered here are,  $1)$ multi-class classifier with softmax labelling, and $2)$ multi-label classifier with sigmoid labelling. 

\subsubsection{Multi-Class Classifier and Softmax Labelling} 
\label{subsubsec:softmax}
Figure \ref{fig:nn_output_e_left} shows the mapping of the labels to the groups when softmax labels are used and Table \ref{tab:softmax_def} provides an example of one way to form the softmax labels.
For this case, every label in the set $\boldsymbol\Theta$ has a length $N_{\textrm{label}}=M^K$, and it identifies a single class that has one member.
As seen in Fig. \ref{fig:nn_output_e_left} and Table \ref{tab:softmax_def}, a class-$n$ has one member which is $\mathbf{x}_n$, and is identified by $\boldsymbol{\theta}_n$.
Obtaining the detector's solution $\hat{\mathbf{x}}$ implies finding one of the $M^K$ classes via the label estimate $\hat{\boldsymbol{\theta}}$.
This is another way of viewing \eqref{eqn:03}, but with a different optimization metric instead of $\ell_2$ metric.
Thus, the QAM-symbol set $\mathcal{D}_{\mathbf{x}}$ is separated into $M^K$ mutually exclusive classes, such that the mapping between the set $\mathcal{D}_{\mathbf{x}}$ and the label set $\boldsymbol\Theta$ is bijective.
As long as a fixed bijective mapping from the set $\mathcal{D}_{\mathbf{x}}$ to the set $\boldsymbol\Theta$ is maintained, these labels may be shuffled.
\begin{table}[!t]
\centering
\renewcommand{\arraystretch}{1}
\caption{An example for softmax label representation for multi-class classifier.}
\label{tab:softmax_def}
\resizebox{\linewidth}{!}{ 
\begin{tabular}{| c | c |}
 \hline
 ${\mathbf{x}}_n = [x_1, x_2]$    & one-hot (softmax) labels $\boldsymbol{\theta}_n$\\
  & \\
 \hline
 $\underbrace{{\mathbf{x}}_{1}}_{\textrm{joint symbol vector}} = \underbrace{[q_1, q_1]}_{\textrm{class }1\textrm{ symbols}}$  &  $\underbrace{\boldsymbol{\theta}_1}_{\textrm{joint label vector}} = \underbrace{[0,0,0,0,0,0,0,0,0,0,0,0,0,0,0,1]}_ {\textrm{class }1\textrm{ label bits}}$  \\ 
  \hline
 ${\mathbf{x}}_{2} = \underbrace{[q_1, q_2]}_{\textrm{class }2\textrm{ symbols}}$  &  $\boldsymbol{\theta}_2 = \underbrace{[0,0,0,0,0,0,0,0,0,0,0,0,0,0,1,0]}_{\textrm{class }2\textrm{ label bits}}$  \\  
  \hline
 ${\mathbf{x}}_{3} = \underbrace{[q_1, q_3]}_{\textrm{class }3\textrm{ symbols}}$   & $\boldsymbol{\theta}_3 = \underbrace{[0,0,0,0,0,0,0,0,0,0,0,0,0,1,0,0]}_{\textrm{class }3\textrm{ label bits}}$  \\
  \hline
 ${\mathbf{x}}_{4} = \underbrace{[q_1, q_4]}_{\textrm{class }4\textrm{ symbols}}$  & $\boldsymbol{\theta}_4 = \underbrace{[0,0,0,0,0,0,0,0,0,0,0,0,1,0,0,0]}_{\textrm{class }4\textrm{ label bits}}$  \\
  \hline
 $\vdots$        & $\vdots$ \\
 \hline
 ${\mathbf{x}}_{13} = \underbrace{[q_4, q_1]}_{\textrm{class }13\textrm{ symbols}}$ & $\boldsymbol{\theta}_{13} = \underbrace{[0,0,0,1,0,0,0,0,0,0,0,0,0,0,0,0]}_{\textrm{class }13\textrm{ label bits}}$  \\ 
  \hline
 ${\mathbf{x}}_{14} = \underbrace{[q_4, q_2]}_{\textrm{class }14\textrm{ symbols}}$  & $\boldsymbol{\theta}_{14} = \underbrace{[0,0,1,0,0,0,0,0,0,0,0,0,0,0,0,0]}_{\textrm{class }14\textrm{ label bits}}$  \\ 
  \hline
 ${\mathbf{x}}_{15} = \underbrace{[q_4, q_3]}_{\textrm{class }15\textrm{ symbols}}$   & $\boldsymbol{\theta}_{15} = \underbrace{[0,1,0,0,0,0,0,0,0,0,0,0,0,0,0,0]}_{\textrm{class }15\textrm{ label bits}}$  \\ 
  \hline
 ${\mathbf{x}}_{16} = \underbrace{[q_4, q_4]}_{\textrm{class }16\textrm{ symbols}}$   & $\boldsymbol{\theta}_{16} = \underbrace{[1,0,0,0,0,0,0,0,0,0,0,0,0,0,0,0]}_{\textrm{class }16\textrm{ label bits}}$  \\
  \hline
\end{tabular}
}
\end{table}

In this case, a loss function known as categorical cross entropy is selected at training.
It is given as
\begin{equation}
\centering
\mathcal{L}(\boldsymbol\Theta,\tilde{\boldsymbol\theta}) =  -\frac{1}{M^K}\sum\limits_{n}^{M^K}\sum\limits_{m}^{M^K} \theta_{nm} \textrm{log}_2(\tilde{\theta}_{m}),
\end{equation}
where $\theta_{nm}$ and $\tilde{\theta}_{m}$ are the $m^{\textrm{th}}$ components of the $n^{\textrm{th}}$ label $\boldsymbol\theta_n$ in the set $\boldsymbol\Theta$ and the NN's output $\tilde{\boldsymbol\theta}$, respectively.
The selected NN's output layer activation function in this case is the softmax function.
If $\mathbf{d}$ is the output from the previous hidden layer, the softmax output is evaluated as
\begin{equation}
\label{eqn:softmax_eqn}
\centering
\tilde{\theta}_m= \frac{\textrm{exp}(d_m)}{\sum\limits_{i=1}^{M^K} \textrm{exp}(d_i)}, \forall m,
\end{equation}
where $\textrm{exp}(\cdot)$ is the exponential function.
It can be seen that \eqref{eqn:softmax_eqn} is evaluated jointly over all the components of the input $\mathbf{d}$.
Further from \eqref{eqn:softmax_eqn}, $\sum\limits_{m=1}^{M^K} \tilde{\theta}_{m} =1$, and $\tilde{\theta}_{m} \geq 0, \forall m$.
Thus the NN's output $\tilde{\boldsymbol\theta}$ for the softmax case defines a probability distribution.
The hard-decision on $\tilde{\boldsymbol\theta}$ to obtain a label estimate $\hat{\boldsymbol{\theta}}$ is given as 
\begin{equation}
\hat{\boldsymbol{\theta}} = \{ m:{\text{arg-max}}\;[\tilde{\theta}_{1},\tilde{\theta}_{2},\cdots,\tilde{\theta}_{M^K}] ; \hat{\theta}_{m}=1; \hat{\theta}_{\bar{m} \neq m }=0  \}.
\end{equation}

\subsubsection{Multi-Label Classifier and Sigmoid Labelling}
\label{subsubsec:sigmoid}
Figure \ref{fig:nn_output_e_right} shows the mapping of labels to the classes when sigmoid labels are used and Table \ref{tab:sigmoid_def} shows an example of one way to form the sigmoid labels.
Every label in the set $\boldsymbol\Theta$ has a length $N_{\textrm{label}}=\text{log}_2(M^K)$ and can be split into $K$ non-overlapping parts, where each part identifies one class.
As seen in Fig. \ref{fig:nn_output_e_right}, each class has one member, which is one of the symbols in the QAM-alphabet $\mathcal{Q}$.
Therefore, the mapping of a part of the label to the class it identifies is bijective.
As long as the bijective mapping between the label set $\boldsymbol\Theta$ and the set $\mathcal{Q}$ holds, the ordering may be shuffled.
Thus, obtaining the detector's solution $\hat{\mathbf{x}}$ implies simultaneously finding $K$ classes.
\begin{table}[!t]
\centering
\renewcommand{\arraystretch}{1}
\caption{An example for sigmoid label representation for multi-label classifier.}
\label{tab:sigmoid_def}
\resizebox{\linewidth}{!}{ 
\begin{tabular}{| c |  c |}
 \hline
 ${\mathbf{x}}_n = [x_1, x_2]$   &  sigmoid labels $\boldsymbol{\theta}_n$ \\
  & \\
 \hline
 $\underbrace{{\mathbf{x}}_{1}}_{\textrm{joint symbol vector}} = [\underbrace{q_1}_{\textrm{class }1 \textrm{ symbol}},\;\; \underbrace{q_1}_{\textrm{class }1 \textrm{ symbol}}]$   & $\underbrace{\boldsymbol{\theta}_1}_ {\textrm{joint label vector}} = [\underbrace{0,0}_{\textrm{class }1 \textrm{ label bits}},\;\; \underbrace{0,0}_{\textrm{class }1 \textrm{ label bits}}]$ \\ 
  \hline
 ${\mathbf{x}}_{2} = [\underbrace{q_1}_{\textrm{class }1\textrm{ symbol} }, \;\; \underbrace{q_2}_{\textrm{class }2 \textrm{ symbol}}]$   & $\boldsymbol{\theta}_2 = [\underbrace{0,0}_{\textrm{class }1\textrm{ label bits}}, \;\;\underbrace{0,1}_{\textrm{class }2 \textrm{ label bits}}]$ \\  
  \hline
 ${\mathbf{x}}_{3} = [\underbrace{q_1}_{\textrm{class }1\textrm{ symbol}}, \;\;\underbrace{q_3}_{\textrm{class }3 \textrm{ symbol}}]$   & $\boldsymbol{\theta}_3 = [\underbrace{0,0}_{\textrm{class }1\textrm{ label bits}},\;\;\underbrace{1,0}_{\textrm{class }3 \textrm{ label bits}}]$ \\
  \hline
 ${\mathbf{x}}_{4} = [\underbrace{q_1}_{\textrm{class }1\textrm{ symbol}}, \;\;\underbrace{q_4}_{\textrm{class }4 \textrm{ symbol}}]$   & $\boldsymbol{\theta}_4 = [\underbrace{0,0}_{\textrm{class }1\textrm{ label bits}},\;\;\underbrace{1,1}_{\textrm{class }4 \textrm{ label bits}}]$ \\
  \hline
 $\vdots$       & $\vdots$ \\
 \hline
 ${\mathbf{x}}_{13} = [\underbrace{q_4}_{\textrm{class }4\textrm{ symbol}},  \;\;\underbrace{q_1}_{\textrm{class }1\textrm{ symbol}}]$   & $\boldsymbol{\theta}_{13} = [\underbrace{1,1}_{\textrm{class }4\textrm{ label bits}}, \;\; \underbrace{0,0}_{\textrm{class }1\textrm{ label bits}}]$ \\ 
  \hline
 ${\mathbf{x}}_{14} =  [\underbrace{q_4}_{\textrm{class }4\textrm{ symbol}},  \;\;\underbrace{q_2}_{\textrm{class }2\textrm{ symbol}}]$   & $\boldsymbol{\theta}_{14} = [\underbrace{1,1}_{\textrm{class }4\textrm{ label bits}}, \;\; \underbrace{0,1}_{\textrm{class }2\textrm{ label bits}}]$ \\ 
  \hline
 ${\mathbf{x}}_{15} = [\underbrace{q_4}_{\textrm{class }4\textrm{ symbol}}, \;\; \underbrace{q_3}_{\textrm{class }3\textrm{ symbol}}]$   & $\boldsymbol{\theta}_{15} = [\underbrace{1,1}_{\textrm{class }4\textrm{ label bits}}, \;\; \underbrace{1,0}_{\textrm{class }3\textrm{ label bits}}]$ \\ 
  \hline
 ${\mathbf{x}}_{16} =  [\underbrace{q_4}_{\textrm{class }4\textrm{ symbol}},  \;\;\underbrace{q_4}_{\textrm{class }4\textrm{ symbol}}]$   & $\boldsymbol{\theta}_{16} = [\underbrace{1,1}_{\textrm{class }4\textrm{ label bits}}, \;\; \underbrace{1,1}_{\textrm{class }4\textrm{ label bits}}]$ \\
  \hline
\end{tabular}
}
\end{table}

A loss function, called binary cross entropy, is considered in this case and is given as
\begin{equation} 
\centering
\begin{split}
 \mathcal{L}(\boldsymbol\Theta,\tilde{\boldsymbol\theta})=-\frac{1}{M^K}\sum\limits_{n}^{M^K} & \sum\limits_{m}^{\textrm{log}_2(M^K)} \left( \theta_{nm} \textrm{log}_2(\tilde{\theta}_{m}) + (1-\theta_{nm}) \textrm{log}_2(  1-\tilde{\theta}_{m})  \right).
 \end{split}
\end{equation}
And the NN's output layer activation function is the sigmoid function.
For an input $\mathbf{d}$ from the previous hidden layer, the sigmoid output is given as
\begin{equation}
\centering 
\tilde{\theta}_m = 1/\left( 1+\textrm{exp}(- d_m )\right), \forall m.
\end{equation}
The hard-decision is evaluated element-wise and is given as
\begin{equation}
\centering 
  \hat{\theta}_{m} =
    \begin{cases}
      0, & \text{if } \tilde{\theta}_{m}\leq 0.5,\\
      1, & \text{otherwise}.\\
    \end{cases}       
\end{equation}

\subsection{Architecture of the NN}
\label{subsec:nn_arch}
Here, two different NN architectures are considered, $1)$ a fully-connected NN (FC-NN), and $2)$ a convolutional NN (CNN).
(See \cite[Fig 5]{park2019wireless} for more details).
A CNN based architecture is chosen over the FC-NN to reduce the number of NN's trainable parameters with subsequent hidden layers \cite{schmidt2017wireless}.
Such a situation may occur when the NN's input is sparse or when the required NN's output is dependent only on a few elements of the NN's input.
For the MUD task here, the above two are not true, since the NN's input $\boldsymbol\Lambda$ is non-sparse and the NN's output $\tilde{\boldsymbol\theta}$ depends on all the available input information.
For performance comparison of different architectures, both FC-NN and CNN are shown here.

\subsubsection{Fully Connected Neural Network}
\label{subsubsec:fcnn}
In the FC-NN, every node in a hidden layer is connected to every other node in the next layer, hence the name fully-connected (FC).
These nodes are the trainable parameters.
Figure \ref{fig:fcnn_exm} is an example of FC-NN.
\begin{table}[t]
\small
\centering
\renewcommand{\arraystretch}{3}
\caption{Parameter setting for FC-NN.}
\label{tab:fcnn}
\scriptsize
\resizebox{1\linewidth}{!}{
\begin{tabular}{| l | l | l |}
 \hline
 $\textrm{}$ & $K=2$ & $K=4$ \\
  \hline
 Architecture & \makecell[l]{ $3$ FC  hidden layers \\ $[201, 351, 263, \overbrace{ N_{\text{label}} }^{\textrm{depends on } \boldsymbol\Theta} ] $ } & \makecell[l]{ $6$ FC hidden layers \\ $[201, 351, 651, 573, 341, 263, N_{\text{label}}]$ }\\
  \hline
   \makecell[l]{Input to get \\ real valued $\boldsymbol\Lambda$ \\ and dimension } &   \multicolumn{2}{c|}{\makecell[l]{$[\mathbf{y}^T, \hat{\mathbf{h}}_{1}^T, \hat{\mathbf{h}}_{2}^T, \cdots,  \hat{\mathbf{h}}_{K}^T, \overbrace{\mathbf{s}_1^T,\mathbf{s}_2^T,\cdots, \mathbf{s}_K^T}^{\textrm{additional}} ]$\\ \\ $\boldsymbol\Lambda$  dimension $( 2( LN_{\textrm{r}} + KLN_{\textrm{r}}  + \underbrace{KL}_{\textrm{additional}} ) \times 1)$} }    \\
  \hline
\end{tabular}
}
\end{table}
Table \ref{tab:fcnn} shows the FC-NN configuration implemented for the MUD task here.
It shows the number of FC hidden layers and the number of nodes in each FC hidden layer for two values of the number of UEs $K$.
As seen in the table, the architecture changes with $K$.
Though not shown in Table \ref{tab:fcnn} and as explained before, each FC layer is followed by a BN layer and then by a ReLU activation layer.
This is a setup similar to the example in Fig. \ref{fig:fcnn_exm}.

The output layer should have the same number of nodes as the label size $N_{\textrm{label}}$.
In Fig. \ref{fig:fcnn_exm} where $N_{\textrm{label}}=2$, a total of $2$ labels can be obtained with the softmax setup as outlined in Section \ref{subsubsec:softmax}.
However, with the sigmoid setup of Section \ref{subsubsec:sigmoid}, a total of $4$ labels can be obtained.
The number of labels is equal to the number of categories that the NN can identify.

The input $\boldsymbol\Lambda$ is a vector, and is the concatenation of the sum-signal $\mathbf{y}$, vectorized channel estimate $\hat{\mathbf{H}}$, and vectorized spread-matrix $\mathbf{S}$.
In Table \ref{tab:fcnn}, the factor $2$ in $\boldsymbol\Lambda$ is due to the complex to real valued conversion in the preprocessing step.
If at the UEs the pilot symbols (used for channel estimation at the AP) are also spread by the spread-vectors, then $\hat{\mathbf{H}}$ is the estimate for the effective channel $\mathbf{H}_{\text{eff}}$, else $\hat{\mathbf{H}}$ is the estimate for the propagation channel  $\mathbf{H}$.
Hence the inclusion of $\mathbf{S}$ in the input $\boldsymbol\Lambda$ may be seen as an additional information. 

\subsubsection{Convolutional Neural Network}
\label{subsubsec:cnn}
\begin{figure}[!t]
\centering
\includegraphics[]{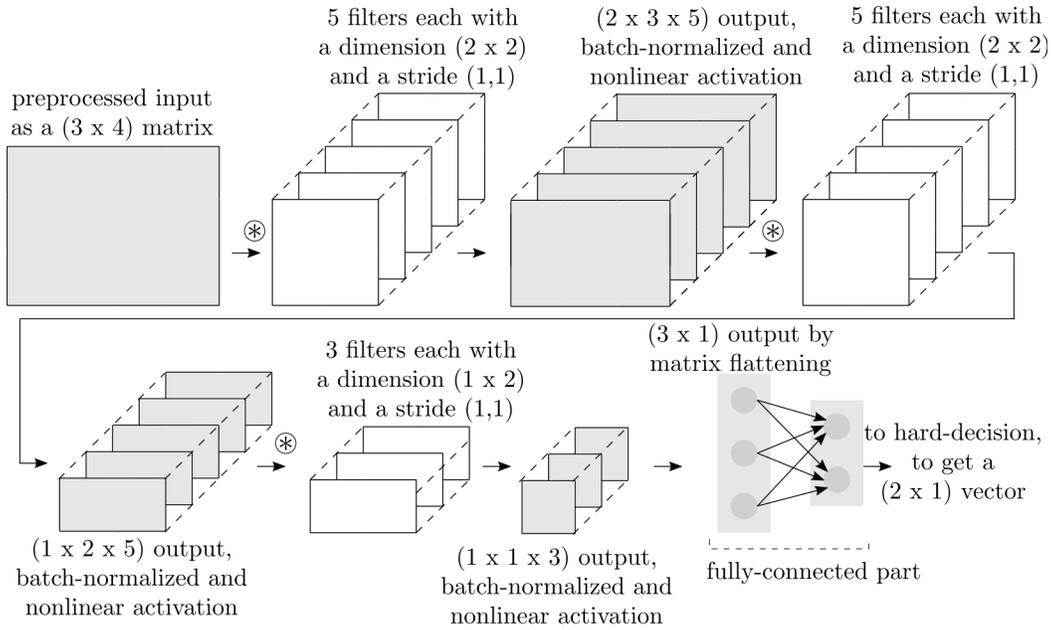} 
\caption{An example of a classifier that is implemented by a $2$D-CNN followed by a FC-NN.}
\label{fig:2dcnn_exm}
\end{figure}

As shown in Fig. \ref{fig:2dcnn_exm}, the architecture of a CNN has several filters.
A filter, also known as a kernel, is a vector or a matrix with trainable coefficients.
The output of each filter is the result of a convolution operation between its input and the filter coefficients.
Hence the name convolutional is used for the NN.
Convolution is performed by sliding a filter over its input.
So in addition to the selection of filter-size, the sliding length, also known as the filter-stride, needs to be set.
Unless there is a zero-padding to the filter's input, the convolution will reduce the filter output's dimension, compared to its input dimension.
This is due to the sliding being restricted to within its input dimensions. 
See \cite{cs231ngithub, zhang2019deep} for more details.

Two types of CNN, namely, $2$D-convolution based CNN, and $1$D-Convolution based CNN, exist.
The main difference between these types is their respective filter-size and filter-stride.
In the former case both the filter dimension and the NN's input $\boldsymbol\Lambda$ are $2$D, with the $2$D-filter sliding over both the length and the breadth of its $2$D-input.
In the latter, both the filter-size and the filter-stride are $1$D, with the $1$D-filter sliding only along the length of its $1$D-input.
Here only $2$D-CNN is considered, while with suitable changes it is straightforward to adapt the same technique for a $1$D-CNN implementation.

Fig. \ref{fig:2dcnn_exm} shows an example of a $2$D-CNN implementation with multiple filters of the different sizes and a FC output layer.
The convolution operation is shown as $\boldsymbol\circledast$.
The NN's input $\boldsymbol\Lambda$ is $2$D and as before, the output of each layer (filter output here) is sequentially sent through a BN layer and a ReLU activation layer.
The output of the activation layer is the input to the convolution filter of the next hidden layer.
As shown in Fig. \ref{fig:2dcnn_exm}, there are multiple filters for a hidden layer, each with the same size and stride.
Since there is no zero-padding (mentioned as valid in Table \ref{tab:2dcnn}) in the hidden layers, the size of the output from each hidden layer is reduced.
As the data passes through the NN towards the output, it is required to select an appropriate number of filters, along with a proper size and stride, to fit the next layer's input.

For the NN to function as a classifier, it's output layer must be a FC implementation.
Since the FC layer takes only $1$D inputs, a flattening layer is used to provide a vectorized input.
In general, when the NN's input $\boldsymbol\Lambda$ is large and the objective is feature extraction, e.g., spatial features here, a pooling layer included after every activation layer or a larger stride-size is used between convolutions.
This will ensure that the input to the subsequent hidden layers is relatively smaller, however it will result in neglecting some intermediate information from some of the components.
So here a stride-size of $1$ is maintained and no pooling is used.

Table \ref{tab:2dcnn} shows the $2$D-CNN configuration used for the MUD task here for two values of the number of UEs $K$.
The architecture changes with $K$.
\begin{table}[t]
\small
\centering
\renewcommand{\arraystretch}{3}
\caption{Parameter setting for $2$D-CNN.}
\label{tab:2dcnn}
\scriptsize
\resizebox{1\linewidth}{!}{
\begin{tabular}{| l | l | l |}
 \hline
 $\textrm{}$ & $K=2$ & $K=4$ \\
  \hline
 architecture & \makecell[l]{ $2$ hidden layers, \\  $65$ filters is each layer, filter size $=(2,2)$, \\ strides $=(1,1)$, padding$=$valid \\ \\  $5$ hidden layers, \\ $37$ filters is each layer, filter size $=(1,2)$, \\ strides $=(1,1)$, padding$=$valid \\ \\ $1$ flatten layer \\  $1$ FC hidden layer with $ \overbrace{ N_{\text{label}} }^{\textrm{depends on } \boldsymbol\Theta}$ units } & \makecell[l]{ $4$ hidden layers, \\ $193$ filters is each layer, filter size $=(2,2)$, \\ strides $=(1,1)$, padding$=$valid \\ \\  $4$ hidden layers, \\ $64$ filters is each layer,  filter size $=(1,2)$, \\ strides $=(1,1)$, padding$=$valid \\ \\ $1$ flatten layer \\ \\ $1$ FC hidden layer with $N_{\text{label}}$ units}\\
  \hline
   \makecell[l]{input to get \\ real valued $\boldsymbol\Lambda$ \\ and dimension } &   \multicolumn{2}{c|}{\makecell[l]{$[\mathbf{y}, \hat{\mathbf{h}}_{1}, \hat{\mathbf{h}}_{2}, \cdots,  \hat{\mathbf{h}}_{K}, \overbrace{\textrm{0-padded }\mathbf{S}}^{\textrm{additional}} ]$. Each  $\mathbf{s}_k \in \mathbf{S}$  is $0$-padded with $ ( N_{\textrm{r}} -1 )L$ zeros. \\ \\ $\boldsymbol\Lambda$  dimension $( 2 LN_{\textrm{r}} \times (K+1  + \overbrace{K}^{\textrm{additional}} ) \times \overbrace{1}^{\textrm{filter dimension}} )$} }    \\
  \hline
\end{tabular}
}
\end{table}
To form a $2$D input, the NN's input $\boldsymbol\Lambda$ is formed by stacking the sum-signal $\mathbf{y}$ along the column of the channel estimates $\hat{\mathbf{H}}$.
If the spread-matrix $\mathbf{S}$ is included into $\boldsymbol\Lambda$, then each column in $\mathbf{S}$ is zero-padded in the preprocessing step to fit the required dimension.

\section{Simulation Results}
\label{sec:results}
In this section, we present the simulation results, and compare the performance of the NN-based schemes with those obtained by ML-based detector.
The modulation is $4$-QAM such that there are $M=4$ symbol per UE, the number of receive antennas $N_{\textrm{r}}=3$, the spread-length $L=2$.
For the spread-matrix $\mathbf{S}$, a Grassmann set is pregenerated when the number of UEs $K=4$, such that correlation among the every two vectors is $\rho=0.577$.
Here, the performance only w.r.t overloading $100\%$ and $200\%$, corresponding to $K=2$ and $K=4$ respectively, is shown.
When the number of UEs $K=2$, the first two SSs are used out of the generated $4$ SSs.

With the channel variance $\sigma^2_{\text{h}}=1/(L N_\text{r})$, and the QAM-constellation $\mathcal{Q}$ normalized as $\frac{1}{M}\sum\limits_{m=1}^{M}(x_m x_m^*)=\textrm{log}_2(M)$, the average per-UE received signal energy (summed over all the $LN_{\textrm{r}}$ REs) has a unit value.
For the test phase noise variance $\sigma^2_{\mathbf{z}}$, the metric $\gamma = 1 / ( LN_{\textrm{r}} \sigma^2_{\mathbf{z}})$ is the average per-UE SNR evaluated per-RE and per-receive antenna.

Fading is independent across both the frequency and the spatial domains for all UEs.
It is assumed that the overall channel $\mathbf{H}$ from all $K$ UEs is perfectly known at the AP at both the training and the testing phases.
Also $\hat{\mathbf{H}}$ is the estimate of the propagation channel $\mathbf{H}$ only.

The spread-matrix $\mathbf{S}$ is not fed into the NN's input $\boldsymbol\Lambda$.
Every complex number in the NN's input $\boldsymbol\Lambda$ is changed to real valued by stacking it's real components next to each other, e.g., $[d_1+jd_2, d_3+jd_4]$ is changed to $[d_1, d_2, d_3, d_4]$.
The NN's learning rate $\alpha=10^{-3}$, and it is trained at a per-RE per-receive antenna training noise variance $\sigma^2_{\mathbf{z}, \text{train}} = -18$ dB.

As mentioned before, the dataset is infinitely large.
Therefore a sampling technique is needed to generate the channel $\mathbf{H}$ and noise $\mathbf{z}$ realizations in the training dataset.
Within an epoch there is only one noise realization and $M^K$ different channel realizations, one each for a member of the set $\mathcal{D}_{\mathbf{x}}$.
For a new epoch, the set $\mathcal{D}_{\mathbf{x}}$ is shuffled, a new noise realization and new set of $M^K$ channel realizations are generated.

The finite dataset is small and a batch-size $N_{\text{batch}}=M^K$ suffices, i.e., $N_{\text{batch}}=16$ when $K=2$ and $N_{\text{batch}}=256$ when $K=4$.
For the testing phase sampling, different noise and channel realizations for each sample of the QAM-symbol set $\mathcal{D}_{\mathbf{x}}$ are considered.
A total of $10^{4}$ noise and per-UE channel realizations are considered in the test phase.

\begin{figure}[t] 
    \centering
    \hspace*{-3mm}   
  \includegraphics[width=.8\textwidth]{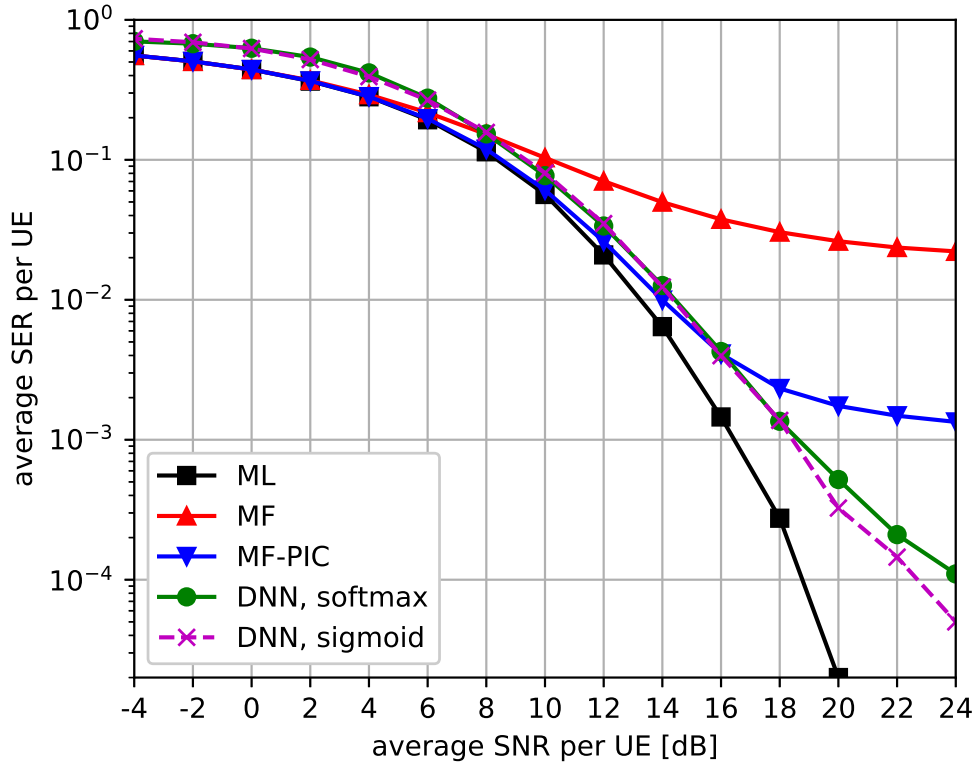} 
  \caption{Comparision of FCNN with conventional detectors. $K=2$, $N_{\text{epochs}}=10^7$.}
  \label{fig:2_ues_b} 
\end{figure}

\begin{figure}[t] 
    \centering
    \hspace*{-3mm}   
  \includegraphics[width=0.8\textwidth]{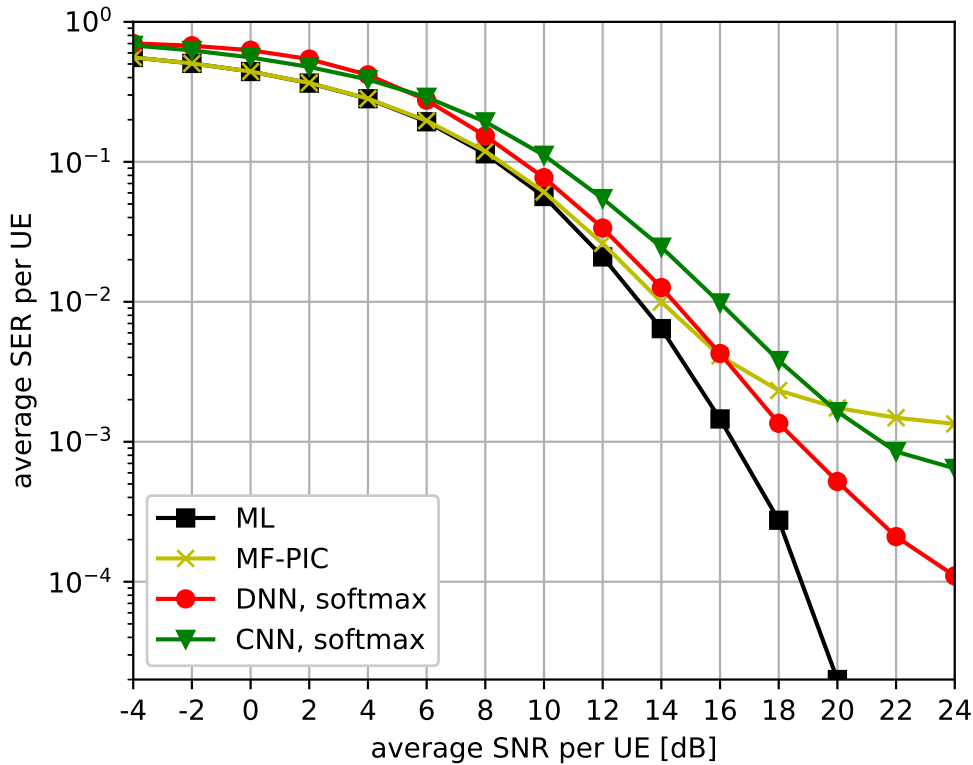} 
  \caption{Comparison of FCNN with $2$D-CNN with softmax labelling. $K=2$, $N_{\text{epoch}}=10^7$.}
  \label{fig:2_ues_c} 
\end{figure}

\begin{figure}[t] 
    \centering
    \hspace*{-3mm}   
  \includegraphics[width=0.8\textwidth]{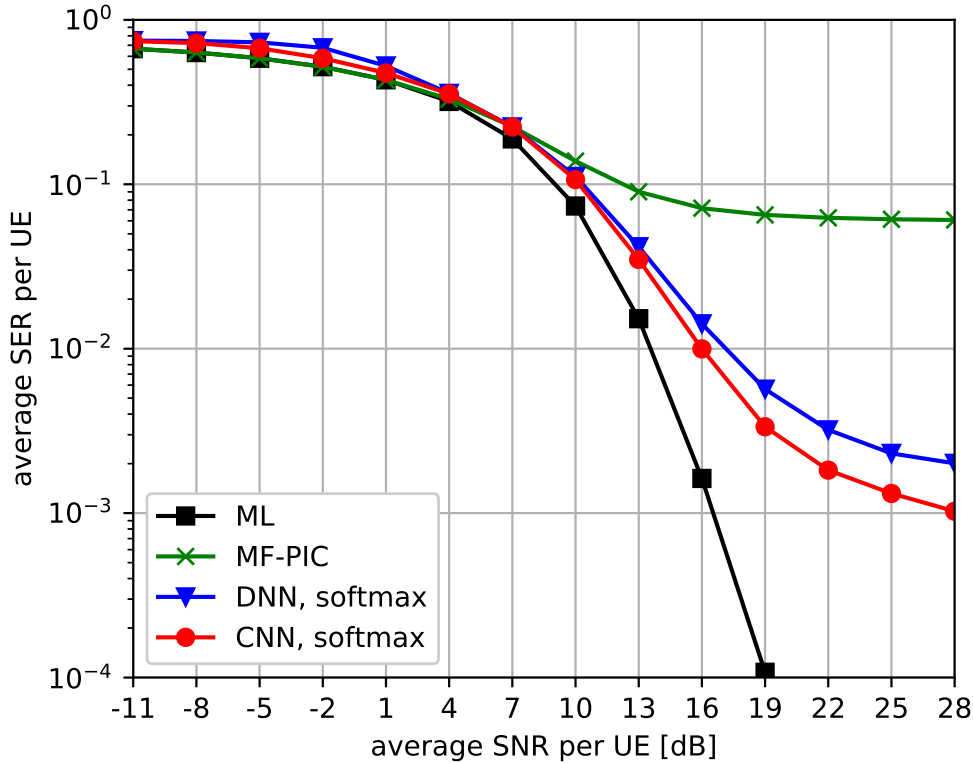} 
  \caption{Comparison of FCNN with $2$D-CNN with softmax labelling. $K=4$, $N_{\text{epoch}}=2\times 10^7$.} 
  \label{fig:4_ues_a} 
\end{figure}

Figures \ref{fig:2_ues_b} - \ref{fig:4_ues_a} show the test phase performance, for various cases.
The average SER per UE versus the average SNR per UE considered over the $LN_{\textrm{r}}$ REs (given as $1 / \sigma^2_{\mathbf{z}} = LN_{\textrm{r}}\gamma$) is plotted.
The SER performance is an indication for the uncoded BER.
The simulation has been implemented in a python based environment where Keras \cite{chollet2015keras} is used for modelling and TensorFlow \cite{tensorflow2015-whitepaper} is used as the backend libraries.

Figure \ref{fig:2_ues_b} shows the FCNN's performance for the two labelling formats when $K=2$ and $N_{\text{epochs}}=10^7$.
In addition, comparison with model based detectors based on MF,  $3$-iteration MF-PIC and ML is shown.
As expected, the ML's performance is the best, while the MF's performance saturates quickly.
At $10 \%$ target-SER, the MF-PIC's plot is about $1$ dB better than both the FCNNs, but at $1 \%$ target-SER this SNR gap is almost negligible.
Beyond $16$ dB, the MF-PIC, due to its interference-limited nature, approaches saturation while the FCNNs have an improved SER.
Clearly, the NN's performance gain is visible from about $1 \%$ target-SER.
For the same amount of training, i.e., $N_{\textrm{epoch}}$, and SNR beyond $18$ dB, the NN's performance with the sigmoid labelling is slightly better than the softmax labelling. 
This may be due to the hyper-parameter tuning.
It is possible to have a completely orthogonal setup when $K=L=2$ where each UE transmits over a non-overlapping RE. 
This is not considered here.

Figure \ref{fig:2_ues_c} compares the SER performance of the FCNN and $2$D-CNN with softmax labelling.
The MF-PIC outperforms the $2$D-CNN for almost the entire range of considered SNR, but the latter has a lower SER floor beyond $20$ dB.
At $10\%$ target-SER, the performance loss of the $2$D-CNN over the FCNN is about $1$ dB, which gradually increases to $2$ dB at $1\%$ target-SER.
The $2$D-CNN operates with lower number of trainable parameters compared to the FCNN, and the $2$D-CNN's performance may be compromised due to this.
A better performance of the $2$D-CNN may be expected by hyper-parameter tuning and/or modifying it's architecture.

Figure \ref{fig:4_ues_a} compares the SER performance of the FCNN and $2$D-CNN with softmax labels for the number of UEs $K=4$.
The number of epochs $N_{\textrm{epoch}}$ is increased to $2\times 10^{7}$.
Unlike the case when $K=2$, the performance gap between the NNs and the ML at low SNR and upto $10\%$ target-SER is almost negligible.
At $10\%$ target-SER, the NNs' performance is better than that obtained by MF-PIC by $2$ dB.
The MF-PIC curve saturates beyond $13$ dB while the NN's saturation is observed beyond $22$ dB.
Though the FCNN and the $2$D-CNN have the same $N_{\textrm{epoch}}$, the $2$D-CNN's performance is better than the FCNN, intuitively due to the hyper-parameter and the architecture selection.

\section{Conclusion}
\label{sec:conclusion}
For a baseband WSMA-NOMA setup, we developed different DL-based schemes for MUD. Two different NN architectures, an FCNN and a 2D-CNN, were considered, and we compared the performance of the proposed schemes with the optimal but complex ML-based schemes. 
The NNs are trained to learn several MUD subtasks such as equalization, combining, slicing, signal reconstruction and interference cancellation.
In addition, the SS to UE assignment, and real-valued to complex-valued demapping is also learnt.
A sampling technique to sweep across the training variables, whose statistics are only known, have also been outlined.
The NNs considered in this paper provide an end-to-end receiver side solution, i.e., the NNs obtain a direct solution to the MUD problem without the individual block outputs as in the case of a conventional receiver.

In the test phase, a trained NN predicts the output from its preprocessed inputs.
Two different NN architectures, an FCNN and a $2$D-CNN, are separately considered for this.
The chosen architectures may not represent an optimal implementation, e.g., optimal values for the hyper-parameters, the number of hidden layers, the nodes per hidden layer, the number of filters, filter-sizes, filter-strides, and so on.
Multiple combinations of various parameters need to be tried before finalizing.

At low and moderate SNRs, our proposed NN-based solutions provide reasonable average SER per UE  close to those obtained by ML, with considerably lower  complexity and no need for closed-form expressions. Moreover, our proposed schemes can be easily adapted for the cases with different channel conditions and transmission parameters.  
At high SNRs, however, the SER plots in a multiuser system saturate due to its interference-limited nature.
For the NN, it could also mean an inaccurate one-shot approximation.
The SER floor may be further reduced by increasing the training epochs, considering a deeper architecture, increasing the number of filters or possibly sweeping over a range of training SNRs.

As we show, with sigmoid labels, the number of output layer nodes reduces considerably, compared to the  cases using softmax labels.
This reduces the number of trainable parameters and will be helpful in the cases where the training set expands, e.g., when there are higher modulation schemes and higher user density.
Since only linear operations and element-wise nonlinear functions are involved in the NN test phase, under some specific time invariant scenarios, an NN is a good substitution for the conventional receiver blocks due to its simplicity in evaluation and low latency prediction.

\bibliographystyle{IEEEtran}  
\bibliography{bib_fileBM}  

\end{document}